\documentclass[a4paper,100pt]{article}
\usepackage{graphicx} 
\usepackage[utf8]{inputenc}
\title{Snowplow Model Predictions for Plasma Temperature in Z pinch Discharges}
\author{Miguel Cárdenas, Alejandro Nettle and Leandro Nú\~{n}ez\\Universidad de Playa Ancha\\Av. Playa Ancha 850, Valparaíso, Chile\\e-mail: miguel.cardenas@upla.cl}

\begin{document}
\date{}

\large
\maketitle
\begin{abstract}

\end{abstract}

Using the general framework of the snowplow model, we introduce the equations for computing plasma temperature in Z pinch discharges. We also present expressions to quantify energy transfers in Z pinch discharges. We then apply this methodology to estimate the temperature and energy transfers of a real prototypical experiment and analyze how the plasma temperature behaves under various modifications of the parameters of the prototypical experiment. Among our discoveries we highlight that the temperature of the plasma in the discharges grows linearly with the initial voltage at the capacitor bank. We summarize this and other findings through a single, fairly compact formula.

\section{Introduction}
The basic theory of how thermonuclear fusion works has long been well established \cite{gamow}. This theory has been corroborated by experiments such as the hydrogen bomb which in turn provides an example of how it is possible to artificially induce uncontrolled thermonuclear fusion. The interesting problem, however, is to achieve controlled thermonuclear fusion in the laboratory \cite{teller}.

The race to achieve controlled thermonuclear fusion in the laboratory has historically taken place through two different practical approaches, magnetic confinement fusion (MCF) and inertial confinement fusion (ICF). A great deal of knowledge has been derived from investigations on both those branches however, the main purpose that motivates these investigations is still pendant. In fact,
ignition and subsequent self-sustainment of fusion is not yet a reality. The most ambitious experiment in the area of MCF is the International Thermonuclear Experimental Reactor (ITER) but it is still under construction. With respect to the field of ICF, the National Ignition Facility (NIF) megajoule laser at the Lawrence Livermore National Laboratory (LLNL) has yielded remarkable results but the research is not yet exhausted \cite{mccall2}.

Research on MCF is essentially based on the exploitation of the phenomenon known as pinch effect \cite{bennett}\cite{tonks}. That is to say, the self-constriction that a current of plasma experiences as a result of coupling with its own magnetic field \cite{jackson}\cite{reitz-milford}\cite{krall}. Based on this principle, a large number of experiments have been designed and built \cite{glasstone}\cite{bishop}\cite{hagler}. Some of them are of the straight discharge tube type or eventually of the toroidal discharge tube type and others are of the tokamak type, like it is for instance the ITER, where the shape of the discharge tube is toroidal but the discharge tube itself is not bare but is surrounded by a complicated arrangement of magnets. 

Straight discharge tube configurations are known as Z pinch devices \cite{post}\cite{kolb}\cite{ryutov}. Currently the 
most successful among the existing Z pinch experiments is the Z Pulsed Power Facility at Sandia National Laboratory. This consists of a huge machine capable of storing 20 MJ of electrostatic energy in its bank of capacitors.

In this work, we focus on the study of discharges in Z pinch devices however, our results are also perfectly extendable to the cases of discharges in toroidal containers, discharges in plasma focus apparatuses \cite{mather}, etc. Our approach to the problem is a mathematical scheme that emulates real experiments within certain limits. Of course, the predictions obtained from this mathematical scheme must be compared with real-world experimental results for validation. In any case, the advantage of having a theoretical basis for simulating real experiments is that the design and obtaining experimental results are facilitated and made less expensive.

The specific theoretical framework with which we approach this research work is the snowplow model \cite{rosenbluth}\cite{cardenas}. Our focus is to identify through this model the factors that positively impact the performance of discharges in Z pinch devices and the factors that negatively impact it. In particular, we are interested in the temperature behavior of the discharges. This topic is particularly important because the temperature of the plasma is the determining factor,\textit{ i.e.} the \textit{conditio sine qua non}, for thermonuclear fusion to occur \cite{lawson}. In this connection, it is worth mentioning that the temperature of the plasma cannot be directly measured in actual Z pinch discharges, but is trivially predicted by the snowplow model \cite{cardenas}.

Before analyzing the behavior of the plasma temperature in Z pinch discharges, we have examined energy transfer in those discharges. For that purpose, we used the concept of efficiency and applied it to accurately quantify energy transfer in Z pinch discharges. We obtained that the efficiency with which the bank of capacitors transfers energy to the plasma is around $51$ percent while the efficiency with which the energy gained by the plasma in that way is actually converted into internal energy of the plasma is around $68$ percent, so the efficiency with which the electrostatic energy initially stored in the bank of capacitors converts into internal energy of the plasma is around $35$ percent.

In relation to temperature, we found out that the temperature of the plasma in Z pinch discharges grows linearly with the initial voltage at the bank of capacitors; the higher the voltage, the higher the temperature of the plasma in the discharge. In turn this implies that the temperature of the plasma in the discharge grows with the square root of the energy initially stored in the bank of capacitors. Of course, this result must be looked at cautiously. A very high voltage at the bank of capacitor could lead to a super-fast discharge, meaning the model we are using becomes invalid. Hence, the temperature predicted by the model in that limiting situation is simply meaningless. If that is the case, a model based on shock heating would  be more appropriate to address the issue. 

Whatever it is, the snowplow model has allowed us to reveal the relationship between the energy content of a Z pinch experiment, say $E$, and the expected plasma temperature, say $k_BT$. We can summarize what we have discovered is common to a wide variety of situations using the simple formula $(k_BT/k_BT_0)=(E/E_0)^n$ where $E_0$ is the energy content of a system used as a referent and $k_BT_0$ is the temperature the plasma reaches in that system. The value of the exponent $n$ is characteristic of the situation under consideration. To date, what we have discovered and reported in this work is that: A proportional enlargement of all the parameters of a Z pinch experiment with exception of its filling density yields $n=0$; a proportional enlargement of the radius of the discharge tube, the capacity of the bank of capacitors and the voltage at the bank of capacitors altogether yields $n=1/3$ \cite{cardenas} and the enlargement only of the voltage at the bank of capacitors yields $n=1/2$. The real validity of this synthesis exercise is strictly as extensive as the real validity of the snowplow model. So, within its range of validity, this synthesis can be used as a criterion for discerning whether a particular enlargement of a reference experiment is really plausible in the real-world.

In Section 2, we explain the fundamentals of the model and present the rules for calculating.
In Section 3, we show our findings and discuss them at length. We present our conclusions in Section 4.

\section{Theory and model}

The Z pinch experimental setup consists of a cylindrical tube made up of an insulating material whose specifications are: inside radius $r_0$, outside radius $R$ and length $l_0$.
Both ends of this tube are sealed up in the following way: one of its ends gets sealed up by means of a metallic lid that serves as one of the electrodes of the apparatus while its other end gets sealed up by means of a metal cover that sorrounds closely the entire outside surface of the tube. The role this metallic cover plays is then twofold, it serves to seal up one of the ends of the tube and it serves also as the second electrode for the discharges to be practised within the tube. 

To really carry out those discharges, the tube has to be, by means of a dedicated gadget, filled with a chosen gas at the desired working pressure $p_0$. After that, a bank of capacitors of total capacity $C_0$ and charged at voltage $V_0$ is connected, by means of a couple of cables, to the electrodes of the apparatus in the following manner: one of those cables is definitely connected to the metallic cover of the tube while the other cable is connected through a spark gap to the other electrode of the tube. Of course, the cabling inside the bank of capacitors, the capacitors of the bank of capacitors themselves, the cables that connect the bank of capacitors to the tube and the spark gap have all together associated an unavoidable inherent inductance whose net value $L_0$ can be measured experimentally. Figure 1 depicts a schematic representation of a Z pinch experimental setup.

\begin{figure}[!h]
\centering
\includegraphics[scale=1]{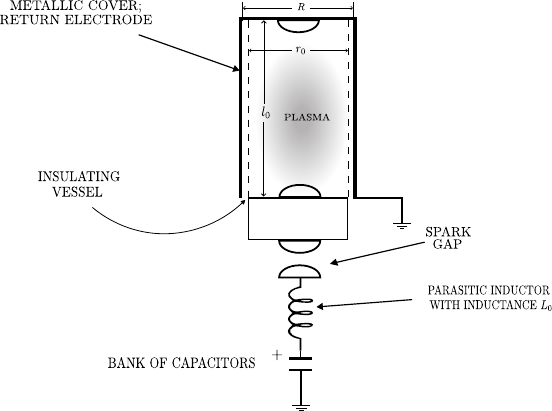}
\caption{Side view drawing of the $z$ pinch experiment.}
\end{figure}

At this stage, it is worth to mention that some care should be taken at the time of choosing $V_0$ and  $p_0$ . Indeed, those variables must be chosen in a way as to guarantee that the electrical breakdown of the working gas contained in the vessel of spatial dimensions $r_0$ and $l_0$  is actually going to take place. Whether or not this will be the case can be easily determined by simple inspection of the Paschen curve for the working gas of the experiment under consideration. This is, we have to check out whether the point $(p_0l_0,V_0)$ lies above that Paschen curve or on the contrary it lies below it \cite{meek}. In the case where the filling gas has been previously ionized, it is not necessary to consider that prevention. 
 
The turning on of the bank of capacitors leads to the electrical breakdown of the gas inside the tube which is followed by an electrical discharge. As time goes on, the discharge evolves to form a sort of column of plasma that, owing to the pinch effect, it gets contracted. Experiments show that this column of plasma, after attaining its maximum contraction, bounces several times before the onset of hydromagnetics instabilities ends up destroying it \cite{glasstone}.  

In this work, we adopt the snowplow model to study numerically discharges of the Z pinch type. However, we will be dealing not exactly with the original set of snowplow equations but instead we will be dealing with a different set of equations derived within the general framework of the snowplow model \cite{cardenas}. Those equations result from taking into account a term that was not considered at the original derivation of the snowplow equations. We refer specifically to the term corresponding to the force outward that the kinetic pressure of the plasma exerts on the current sheath.

Thus, the computations and the reasonings we will display along this article do not rest on the original snowplow (SP) equations but rather they rest on the modified snowplow (MSP) equations. 

The model that give rise to the MSP equations is based on a number of assumptions that, in any case, tend to idealize the performance of the plasma toward the pinch formation and the ultimately purpose  of heating up the plasma inside the cylindrical vessel. 
In that sense, the temperature of the plasma as predicted from a particular simulation within the snowplow model may be regarded as an upper bound for the temperature the plasma, whatever it is, may reach in the corresponding actual experiment.   

The assumptions from which the MSP equations result can be summarized as follows:

\begin{enumerate}
\item The gas inside the tube is assumed to be ionized before the bank of capacitors is switched on and the conductivity of the gas is infinite.
\item Owing to the skin effect, the flowing plasma sets up along an infinitesimally thin cylindrical shell that prevails during the discharge \cite{spitzer}. As time goes on, this shell contracts toward the axis of the tube and at the same time it sweeps up all the particles it encounters. In this manner, the space between the current sheath and the wall of the tube is a vacuum, so that the plasma 
enclosed by the current sheath is kept, during compression, thermally isolated. Hence, this process may be regarded as a case of adiabatic compression. 
\item The particles enclosed by the sheath of current undergo inelastic collisions with that moving shell. In this way, those particles gain kinetic energy.
\item The velocities acquired by the particles after colliding with the shell randomnize rapidly in a fashion that the kinetic energy gained by all the particles transforms genuinely to internal energy of the gas. If this were not the case, we would have instead beams of particles moving coherently and the plasma would not raise its temperature at all.
\item We compute the temperature of the plasma at the time $t_p$ of pinching by identifying the kinetic energy gained by all the particles of the plasma during the interval $0\to t_p$ with the internal energy of the plasma.
Of course, prior to the discharge, the gas inside the tube has both a ground internal energy and correspondingly a ground temperature. Despite this, along our algebraic manipulations, we will ignore those terms because they are really irrelevant to determine the final state of the plasma but not only for that, we will ignore them also because in that way the algebra gets simpler.
\item Losses due to bremsstrahlung radiation are ignored.
\item Instabilities like hydromagnetics instabilities and that sort are not considered.
\item Any ohmic resistance presents in the cabling, the tube, etc is ignored
\item The bank of capacitors has a total, constant, capacity $C_0$. Here we have assumed that neither the cabling, the spark gap nor the tube plus the gas behaves as a capacitor. This is, everyone of them has nominal capacity $C_i=0$ where $i$ stands for the different pieces just mentioned. Alternatively, if in practice that is not the case, \textit{i.e.} some $C_i\neq0$, then we absorb that $C_i$ into $C_0$.
\item The cabling that connects the various capacitors inside the bank of capacitors, the cables that connect the bank of capacitors with the discharges tube and the spark gap contribute all them to the inherent inductance of the system. The piling up of those contributions is equivalent to have a unique inductor of constant inductance with value $L_0$.
\item Given the configuration of the tube, the inductance of the apparatus changes as the radius of the current shell $r$ changes. To be specific, that part of the system behaves as a coaxial cable where the current sheath corresponds to the central conductor and the metallic cover of the tube plays the role of the return path. 
\item The value of the time dependent total inductance $L(t)$ of the Z-pinch system decomposes as 

\begin{equation}
L(t)=L_0+\left(\frac{\mu_0l_0}{2\pi}\right)\,\ln\left(\frac{r_0}{r}\right)
\end{equation}
where we have assumed that the wall of the insulating tube is very thin, $r_0\approx R$, so that we wrote down in the argument of the logarithm $r_0$ instead of $R$.

N.B. When $l_0$ is expressed in centimeters the term $(\mu_0l_0/2\pi)$ equals $(2\times 10^{-9})\,l_0$ henrys.
\item From the electrical point of view, the whole system consists of a source of energy -a bank of capacitors- of initial voltage $V_0$ and constant capacity $C_0$ that discharges on an inductor with inductance $L(t)$. See Figure 2.
\end{enumerate}

\begin{figure}[!h]
\centering
\includegraphics[scale=.2]{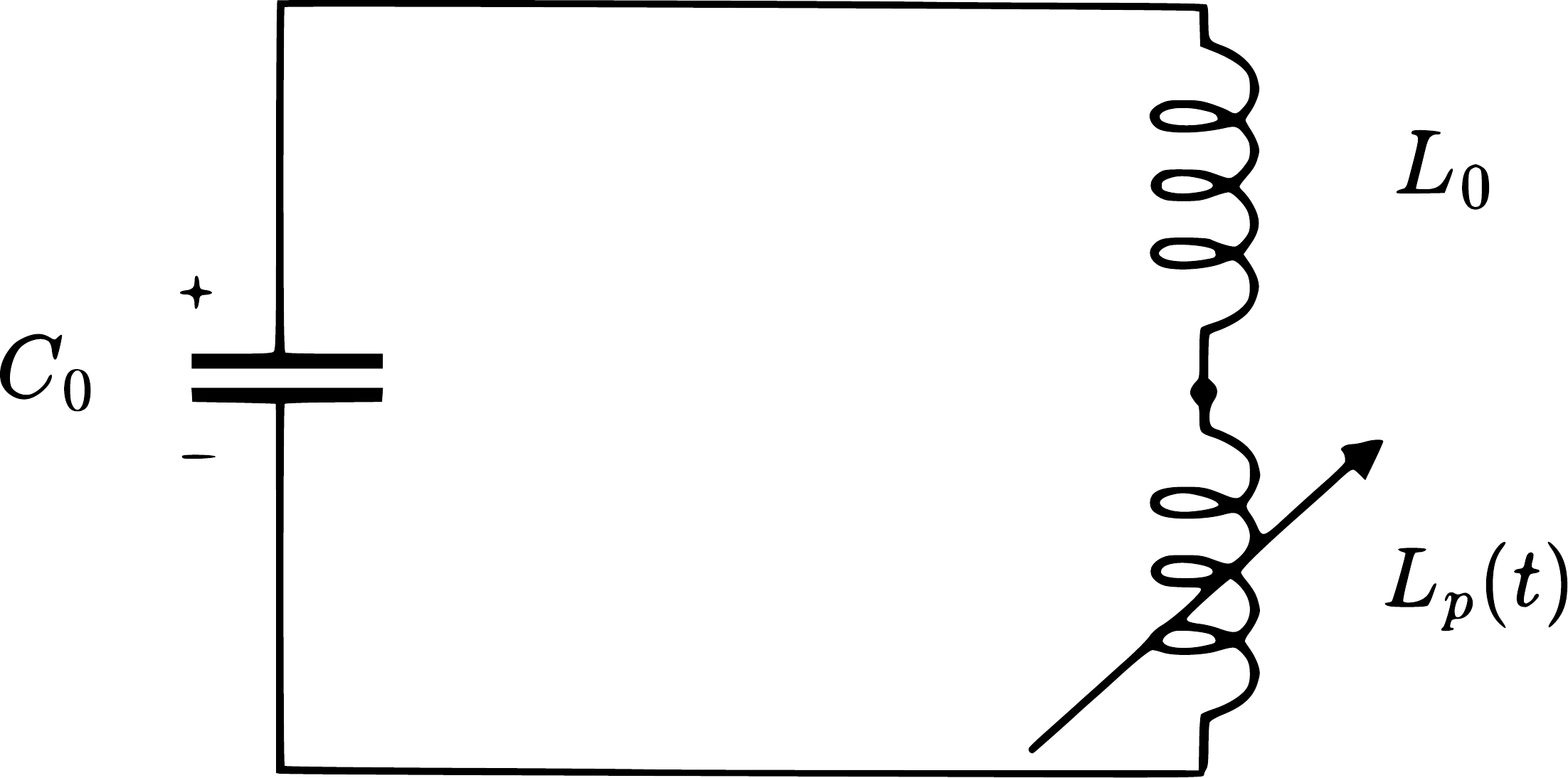}
\caption{Equivalent electrical circuit of the $z$ pinch experiment. Here $L_p(t)$ stands for $L_p(t)=\left(\mu_0 l_0/2\pi\right)\ln\left[r_0/r(t)\right]$.}
\end{figure}

In experiments on Z-pinch discharges, the radius of the plasma column as a function of time $r(t)$ can be reconstructed on the basis of the experimental data for the electrical signals, \textit{i.e.} voltage and current, across the plasma column. Other physical quantities like for instance the temperature of the plasma at pinching are more elusive to measurement. However for the specific case of the temperature, it would be plausible to devise a probe for determining
whether the discharge is suitable or not for heating the plasma to the temperature of interest. Of course, such a probe is intended to be valuable for the case where the temperature of the plasma is assumed to be located far below the range of thermonuclear interest. On the contrary, if we assume \textit{a priori} that the temperature of the plasma falls within the range of thermonuclear interest, then we would like instead to look for the emission of neutrons of thermonuclear origin, \textit{i.e.} we would focus on the detection of isotropically distributed neutron flashes \cite{anderson}\cite{hai}\cite{mc call}.

All the practical limitations for measuring the temperature of the plasma in actual experiments can be somehow circumvented by
using, as a first approximation to deal with them, the MSP equations derived within the framework of the snowplow model. Indeed, the MSP equations puts at our disposal all the physical quantities. Those quantities result straightforwardly from solving numerically two coupled nonlinear integro-differential equations for the radius of the plasma sheath and the current flowing through it as a function of time.

In sections 2.1, 2.2 and 2.3 below, we present the rules for computing: the dynamics of Z pinch discharges (Section 2.1), the internal energy and temperature attained by the plasma in Z pinch discharges (Section 2.2) and energy transfer in Z pinch discharges (Section 2.3). 

\subsection{Dynamics}
Before to introduce the MSP equations for the dynamical variables, we warn that the time $t$, the instantaneous radius of the plasma sheath $r(t)$ and the current flowing through it $I(t)$, as they will appear in those equations, are now dimensionless variables.
Those dimensionless variables result from carrying out the following normalization on the corresponding physical quantities (\textit{i.e.} the normalization on the quantities originally expressed in physical units):
\begin{enumerate}
\item The time $t$ is measured in units of $\sqrt{L_0C_0}$.
\item The instantaneous radius of the plasma sheath $r$ is measured in units of $r_0$. 
\item The current flowing through the plasma sheath $I$ is measured in units of $V_0\,\sqrt{C_0/L_0}$.
\end{enumerate} 
Within the framework of this particular normalization of the dynamical variables, the MSP equations read \cite{cardenas}

\begin{equation}
\frac{d^2r}{dt^2}=\frac{6r^2 (dr/dt)^2-3\alpha^2 I^2-4\int_0^t (1/r)(dr/dt')^3 dt'}{3r(1-r^2)}
\end{equation}
and

\begin{equation}
\frac{dI}{dt}=\frac{1-\int_0^tIdt'+\beta (I/r)(dr/dt)}{1-\beta\ln{(r)}}
\end{equation}
where the dimensionless parameters $\alpha$ and $\beta$ are given by

\begin{equation}
\alpha=\sqrt{\frac{\mu_0 C_0^2  V_0^2}{4\pi^2  r_0^4\rho_0 }}
\end{equation}
and 
\begin{equation}
\beta=\left(\frac{\mu_0 l_0}{2\pi L_0}\right)
\end{equation}
whereas $\rho_0$ stands for the density of the filling gas. Alternatively, that density can be written as $\rho_0=mn_0$ where $m$ represents the mass of the nuclei (or almost equivalently the mass of the atoms for the case of a monoatomic gas) and $n_0$ is the atoms number density.

The initial conditions for the snowplow equations displayed above are:
\begin{equation}
r(0)=1,
\end{equation}

\begin{equation}
I(0)=0,
\end{equation}

\begin{equation}
\left(\frac{dr}{dt}\right)_{t=0}=0,
\end{equation}

\begin{equation}
\left(\frac{d^2r)}{dt^2}\right)_{t=0}=-\left(\frac{\alpha}{\sqrt{3}}\right)
\end{equation}
and

\begin{equation}
\left(\frac{dI}{dt}\right)_{t=0}=1.
\end{equation}

\subsection{Internal energy and temperature}
Once we solve the MSP equations, we are in a position to compute the internal energy $U$ of the gas at time $t_p$. The explicit expression for the internal energy is \cite{cardenas}

\begin{equation}
U=\left(\frac{\beta}{\alpha^2}\right)E_0\int_0^{t_p}\left[-\frac{1}{r}\left(\frac{dr}{dt'}\right)^3\right]dt'
\end{equation}
where $E_0$ corresponds to the electrostatic energy initially stored in the bank of capacitors. This is 
\begin{equation}
E_0=\frac{1}{2}C_0V_0^2.
\end{equation} 
The temperature, in units of energy, at pinching is obtained as 
\begin{equation}
k_BT=\frac{2}{3}\left(\frac{U}{N_0}\right)
\end{equation}
where $N_0$ stands for the number of particles of the gas, this is the number of ions plus the number of electrons. Normally,  we will ignore the electrons in the term $N_0$. Although, this procedure may lead to an inaccuracy in the computation of temperature, the order of magnitude of the so computed temperature is still properly captured. In this connection, it is worth to emphasize here that the estimation of the order of magnitude of the temperature is what we are really focused in.

In closing this subject, we write down the formula for the temperature of the plasma at pinching as a functional of the solution of the MSP equations for $r(t)$ and $I(t)$. 
The formula follows from replacing Equation (11) into Equation (13), so

\begin{equation}
k_BT=\frac{2}{3}\left(\frac{\beta}{\alpha^2}\right)\left(\frac{E_0}{N_0}\right)\int_0^{t_p}\left[-\frac{1}{r}\left(\frac{dr}{dt'}\right)^3\right]dt'
\end{equation}
Clearly, the physical and geometric parameters of the system are summarized in the factor before the integral. Moreover, the parameters $r_0$, $C_0$, $L_0$ and $V_0$ enter that formula also implicitly through the normalization of $r$, $t$ and $I$. The value of the integral is determined by the curve $r(t)$ which in turn is coupled through the MSP equations with the curve $I(t)$. Even more, the particular values taken by the dimensionless parameters $\alpha$ and $\beta$ define the specific form of the MSP equations and in that manner, they also contribute to shaping the curves $r(t)$ and $I(t)$.

Equation(14) can also be written down in a more fashionable way as
\begin{equation}
k_BT=\frac{2}{3}m\left(\frac{r_0^2}{L_0C_0}\right)\int_0^{t_p}\left[-\frac{1}{r}\left(\frac{dr}{dt'}\right)^3\right]dt'.
\end{equation}
Notice that now the term inside the parentheses before the integral is ultimately purely geometric whereas both the integrand and the integral itself are dimensionless.

Finally, for the purpose of comparing this estimation of $k_BT$ with results of actual experiments, we introduce yet another version of $k_BT$. This is 
\begin{equation}
k_BT=\frac{2}{3}m\int_0^{t_p}\left[-\frac{1}{r}\left(\frac{dr}{dt'}\right)^3\right]dt'
\end{equation}
where the integrand is now no longer dimensionless but it is expressed in actual physical units. 

This formula has great practical relevance because the curves $r(t)$ and $(dr/dt)$ constitute part of the data normally obtained from experiments in Z pinch discharges. For a complete description on how those curves may really be extracted from electrical measurements, you might consult reference \cite{glasstone}, p. 247 \textit{et seq}.

Thus, the snowplow model estimation for $k_BT$ can be carried out by replacing the experimental curves $r(t)$ and $(dr/dt)$ in Equation (16), in a way that the resulting $k_BT$ can be clearly compared with the, if any, direct measurement of $k_BT$.

\subsection{Energy transfer and efficiency}
In regards to the power the bank of capacitors delivers to the inductor with inductance $L$, we begin by writing down the Kirchhoff loop rule for the schematic (see Figure 2) that, from an electrical point of view, captures the essence of a Z pinch experiment. 

The Kirchhoff loop law for that circuit yields the equation
\begin{equation}
V+\mathcal{E}=0
\end{equation}
where $V$ is the voltage at the bank of capacitors and $\mathcal{E}$ is the induced electromotive force (emf) accros the inductor. In turn, $\mathcal{E}$ is given, in terms of the flux of magnetic field through the inductor 
\begin{equation}
\Phi=LI
\end{equation}
 by the expression
\begin{equation}
\mathcal{E}=-\left(\frac{d\Phi}{dt}\right),
\end{equation}
therefore
\begin{equation}
V=\left(\frac{d\Phi}{dt}\right).
\end{equation}

The power delivered by the bank of capacitors $P_d$ is obtained as the product of the voltage $V$ times the current flowing in the circuit $I$. After developing Equation (20) that product converts into

\begin{equation}
P_d=\left(\frac{dL}{dt}\right)I^2+LI\left(\frac{dI}{dt}\right).
\end{equation}
Now well, $P_d$ splits up into two pieces. One of those pieces, $P_i$, transfers directly into plasma whereas the other piece, $P_s$,
is diverted. It is used to provide and store energy in the inductor with inductance $L$. The energy stored in the inductor at time $t$ is given by
\begin{equation}
E_s=\frac{1}{2}LI^2
\end{equation}
therefore the rate at which the energy enters the inductor $P_s$ (\textit{i.e.} the instantaneous power transferred to the inductance) is obtained as
\begin{equation}
P_s=\frac{d}{dt}\left(\frac{1}{2}LI^2\right)
\end{equation}
that after performing the time derivative becomes
\begin{equation}
P_s=\frac{1}{2}\left(\frac{dL}{dt}\right)I^2+LI\left(\frac{dI}{dt}\right).
\end{equation}
Then the piece of power $P_i$ is simply computed as 
\begin{equation}
P_i=P_d-P_s
\end{equation}
which, after substituting Equation (21) and Equation (24) into its right-hand side, becomes
\begin{equation}
P_i=\frac{1}{2}\left(\frac{dL}{dt}\right)I^2.
\end{equation}

Once we know the expressions for the various pieces of power outlined above, we can compute the corresponding amounts of energy transferred during a given interval of time. In particular, we are interested in the extent of the energies transferred within the interval of time that goes from the beginning of the discharge $t=0$ to the time when the first pinch occurs, $t=t_p$.

The specific expression for the energy delivered by the bank of capacitors to the load is
\begin{equation}
E_d=\int_0^{t_p}\left[\left(\frac{dL}{dt}\right)I^2+LI\left(\frac{dI}{dt}\right)\right]dt
\end{equation}
whereas the expression for the energy that enters the plasma is

\begin{equation}
E_i=\int_0^{t_p}\left[\frac{1}{2}\left(\frac{dL}{dt}\right)I^2\right] dt.
\end{equation}

These formulas can also be worked out further to cast them as

\begin{equation}
E_d=-2 E_0\left[\beta\int_0^{t_p}\frac{1}{r}\left(\frac{dr}{dt'}\right)I^2dt'-\int_0^{t_p}\left(1-\beta\ln{(r)}\right)I\left(\frac{dI}{dt'}\right)dt'\right]
\end{equation}
and

\begin{equation}
E_i=-\beta E_0 \int_0^{t_p}\frac{1}{r}\left(\frac{dr}{dt'}\right)I^2dt'
\end{equation}
where, we recall, $t$, $r$ and $I$ are dimensionless variables. 

\section{Results and Discussion}

In this section, we revisit the relationship between the size of a Z pinch device and its performance. In Section 3.1, we review the temperature reached by the plasma during discharges. In Section 3.2, we analyze energy transfers during discharges. We close with Section 3.3 where we sketch a procedure that in principle may serve to improve the performance of discharges in Z pinch devices.

\subsection{Temperature}    

In our previous work \cite{cardenas}, we considered a Z pinch system whose set of physical specifications we refresh here for convenience.

\begin{enumerate}
\item Mass of the Atoms of the Filling Gas: $m$,
\item Filling Gas Density: $\rho_0$, 
\item Tube Radius: $r_0$, 
\item Tube Length: $l_0$, 
\item Inherent Inductance (\textit{i.e.} Parasitic Inductance) of the Bank of Capacitors: $L_0$, 
\item Total Capacity of the Bank of Capacitors: $C_0$

and

\item Initial Voltage of the Bank of Capacitors: $V_0$.
\end{enumerate}
In the referred work, we analyzed the behavior of a Z pinch device when subject to the following transformation:
\begin{enumerate}
\item $ m\to m$,
\item $\rho_0\to \rho_0$, 
\item $r_0\to\lambda r_0$, 
\item $l_0\to \lambda l_0$, 
\item $L_0\to \lambda L_0$, 
\item $C_0\to \lambda C_0$

and

\item $V_0\to \lambda V_0$
\end{enumerate}
where $\lambda>1$.

This particular scale  transformation increases by the same factor $\lambda$ the radius and the length of the discharge tube, among other parameters of the Z pinch device, hence the so enlarged discharge tube is geometrically similar to the non enlarged discharge tube. On the other hand, the dimensionless parameters $\alpha$ and $\beta$ remain invariant under the complete tranformation. Therefore, the enlarged system and the non enlarged system share with each other kinematic similarity too. In addition, the energy stored in the bank of capacitors scales as $E_0\to\lambda^3\,E_0$ and similarly does the number of atoms within the tube, $N_0\to\lambda^3\,N_0$ so that, according to Equation (14), the enlarged system and the non enlarged system achieve the same temperature in the discharges \cite{cardenas}.

Here, it is worth to point out that a transformation restricted only to the enlargement of the lengths $r_0$ and $l_0$  is not desirable at all. In effect, such a transformation preserves geometric similarity but harmfully removes kinematics similarity. 

To convey it in other terms, we may affirm that if the actual tube is very long and it is also very large then it will contain many particles, so that the electric field crossing the tube, namely $(V_0/l_0)$, may become -in some sense-  unable to generate a dynamics suitable for heating the plasma.

\subsection{Energy transfers}

A subject worth considering concerns the efficiency of the discharges in Z pinch apparatuses. 
To illustrate the point, let us go ahead with the full transformation we are analyzing. Within that context, the various energies transferred during the interval of time $0\to t_p$ grow significantly, due to the enlargement of the system, in the manner we explain below.
According to Equation (11) the internal energy in the enlarged system grows as $\lambda^3$ with respect to the internal energy in the non enlarged system, this is
\begin{equation}
U(t)\to \lambda^3U(t).
\end{equation}
According to Equation (29) the energy delivered by the bank of capacitors in the enlarged system scales also as  
\begin{equation}
E_d(t)\to \lambda^3 E_d(t).
\end{equation}
Finally, according to Equation (30) the energy that enters the plasma obeys the same rule of scale, this is
\begin{equation}
E_i(t)\to \lambda^3E_i(t)
\end{equation}

In spite of these rules of scaling for the different energies, the efficiency at which any of these energies is transferred in the enlarged system is exactly the same as the efficiency at which the corresponding energy is transferred in the non enlarged system.

To be more precise, if we define -percentually speaking- the efficiency at which the bank of capacitors delivers energy to the plasma as
\begin{equation}
\eta_{pl}=\left(\frac{E_i}{E_d}\right)\times 100
\end{equation}
and analogously, we define the efficiency at which that energy transforms to internal energy of the plasma as
\begin{equation}
\eta_{th}=\left(\frac{ U}{ E_i}\right)\times 100
\end{equation} 
then, from the rules of scale for the different energies displayed above, follows immediately that the values of $\eta_{pl}$ and $\eta_{th}$  remain invariant. This is, 
the value of $\eta_{pl}$ for 
the enlarged system is the same as the value of $\eta_{pl}$ for the non scaled system and the value of $\eta_{th}$
for the enlarged  system is the same as the value of $\eta_{th}$ for the non scaled system.
In synthesis, we can affirm that any two different experiments that share geometrical and kinematic similarity must yield the same values of their corresponding $\eta_{pl}$ and $\eta_{th}$.

The useful expressions to evaluate $\eta_{pl}$ is obtained by dividing Equation (30) by Equation (29), this is

\begin{equation}
\eta_{pl}=\frac{1}{2} \left[\beta\int_0^{t_p}\frac{1}{r}\left(\frac{dr}{dt'}\right)I^2dt'\right]\times 100
\Bigg/\left[\beta\int_0^{t_p}\frac{1}{r}\left(\frac{dr}{dt'}\right)I^2dt'
-\int_0^{t_p}\left(1-\beta\ln{(r)}\right)I\left(\frac{dI}{dt'}\right)dt'\right]
\end{equation}
whereas the useful expression to compute $\eta_{th}$ is obtained by dividing Equation (11) by Equation (30) which gives

\begin{equation}
\eta_{th}=\frac{1}{\alpha^2}\left[\int_0^{t_p}\frac{1}{r}\left(\frac{dr}{dt'}\right)^3dt'\right]\times 100
\Bigg/\left[\int_0^{t_p}\frac{1}{r}\left(\frac{dr}{dt'}\right)I^2dt'\right].
\end{equation}

Now well, the application of the techniques of the snowplow model to the study of a number of actual experiments throws repeatedly out the result
\begin{equation}
\eta_{pl}\approx 51\,\%
\end{equation}
and
\begin{equation}
\eta_{th}\approx 68\,\%,
\end{equation}
so that we may in broad terms state that the efficiency,
\begin{equation}
\eta=\eta_{pl}\times\eta_{th},
\end{equation} 
to which the energy of the bank of capacitors transforms to internal energy of the plasma is $\eta \approx 35\,\%$.

Of course, the characteristic value taken by $\eta_{pl}$ in Z pinch discharges is not a surprise. A similar behavior is of common ocurrence in many electric circuits concerning the discharge of a source of voltage across a load impedance, \textit{e.g.} configurations for which the power transfer between source and load is maximal may however develop a poor efficiency and in conversed way, configurations where the power transfer between source and load is highly efficient may yet give rise to only modest power transfer between source and load.

The characteristic value of $\eta_{th}$ is also quite plausible since there are many examples of processes in which mechanical energy only partially converts to heat, \textit{vis-\`a-vis} internal energy.

\subsection{Hypothetically better performing Z pinch devices}

One of the main challenges of research in Z pinch discharges is to make a suitable device able to heat its working gas to temperatures within the range of thermonuclear interest. In this connection, we have already shown, on the basis of the snowplow model, that the temperatures achieved by actual discharges in a variety of Z pinch devices are on the order of the dozens of $eV$ only. We have also demonstrated that a proportional scaling of the parameters of a given Z pinch device will not result in increased temperatures. Nonetheless, there exists hypothetically an outcome to the issue.
To make the point, let us focus on a successful Z pinch discharge. For instance, we may consider the typical Z pinch experiment we analyzed in a previous work \cite{cardenas}. 

In the corresponding experiment, the tube of about $3\, L$ volume is filled with helium gas at  $1\, mm$ of mercury pressure approximately. The electrostatic energy stored in the bank of capacitors is $E_0\approx 5760\,[J]$
and the potential difference between the anode and cathode is $V_0\approx 12\,[kV]$, in a sort that the electric field applied to the gas $(V_0/l_0)$ is on the order of a few hundred volts per centimeter. Finally, the parasitic inductance is $L_0\approx 30\,[nH]$.
The simulation, using the MSP equations, of discharges in this experiment shows that the temperature of the plasma at pinching is $k_BT_0\approx 32\, eV$. In order to improve this result, we try an action on the voltage of the bank of capacitors. This is, we keep the experimental setup untouched but we rise the voltage of the bank of capacitors from the value $V_0$ to the value $\lambda\, V_0$. In this manner,  the energy stored in the bank of capacitors now, say $E$, satisfies the scaling rule

\begin{equation}
E=\lambda^2\,E_0
\end{equation}
but not only that, the dimensionless parameter $\alpha$ and the unit of current used in normalization $I_0=V_0\,\sqrt{C_0/L_0}$ scale as

\begin{equation}
\alpha\to\lambda\,\alpha
\end{equation}
and

\begin{equation}
I_0\to\lambda\,I_0
\end{equation}
in a way that the factor $\lambda$ does enter Equation (2) and Equation (3).
 
To give us an idea of how the $V_0\to\lambda\,V_0$ transformation works, we consider the case in which $\lambda=4$, so that the electric field applied to the gas increases from the value

\begin{equation}
(V_0/l_0)\approx 330\, [V/cm]
\end{equation}
to the value

\begin{equation}
(4\,V_0/l_0)\approx 1320\, [V/cm].
\end{equation}

With regard to the dynamics of discharges in these two configurations, we draw together, in Figure 3, the curve $r(t)$ for both situations, the case in which the voltage of the bank of capacitors is $V_0$ and the case in which that voltage is $4\,V_0$. Each of these curves is plotted for clarity in the interval of time that goes from $t=0$ to the time when its corresponding first pinch occurs.
Of course, the curves $r(t)$ result from solving the dimensionless MSP equations but they are presented here with their own physical units, \textit{i.e.} $r$ is expressed in centimeters and $t$ is expressed in microseconds.
It is apparent that the higher the voltage at the bank of capacitors, the faster the dynamics of the current sheath.

\begin{figure}[!h]
\centering
\includegraphics[scale=.6]{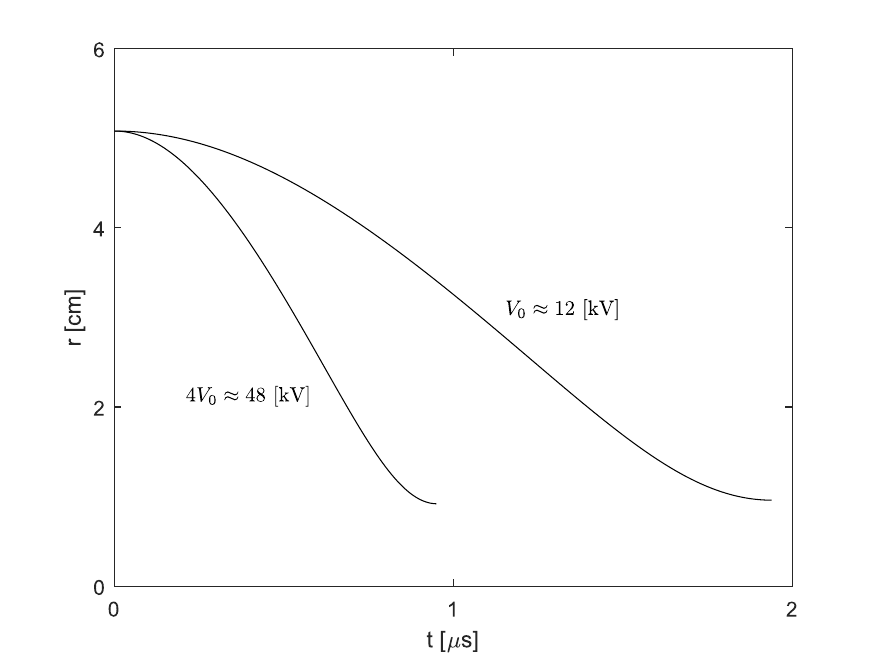}
\caption{Radius of the current sheath $r$ versus time $t$ for the cases $V_0\approx 12$~[kV] and $4V_0\approx 48$~[kV].}
\end{figure}
\pagebreak
In what concerns temperature, we already quoted that for the case in which $V_0\approx 12\,[kV]$, the temperature the plasma reaches at pinching is $k_BT_0\approx 32\,eV$. However, in a not enterely predictable way, when the voltage of the bank of capacitors increases from $V_0\approx 12\,[kV]$ to $4\,V_0\approx 48\,[kV]$, the temperature of the plasma at pinching rises from
$k_BT_0\approx 32\,eV$ to $k_BT\approx 142\,eV$. 

To reveal the path that leads to this outcome, we resort to Equation (16)
for the temperature.
In Figure 4, we show the integrand of Equation (16), namely

\begin{equation}
f(t)=\frac{2}{3}m\left[-\frac{1}{r}\left(\frac{dr}{dt'}\right)^3\right]
\end{equation}
for the two cases we are dealing with. So, temperature quantitatively matches the area under the curve $f(t)$.

\begin{figure}[!h]
\centering
\includegraphics[scale=.6]{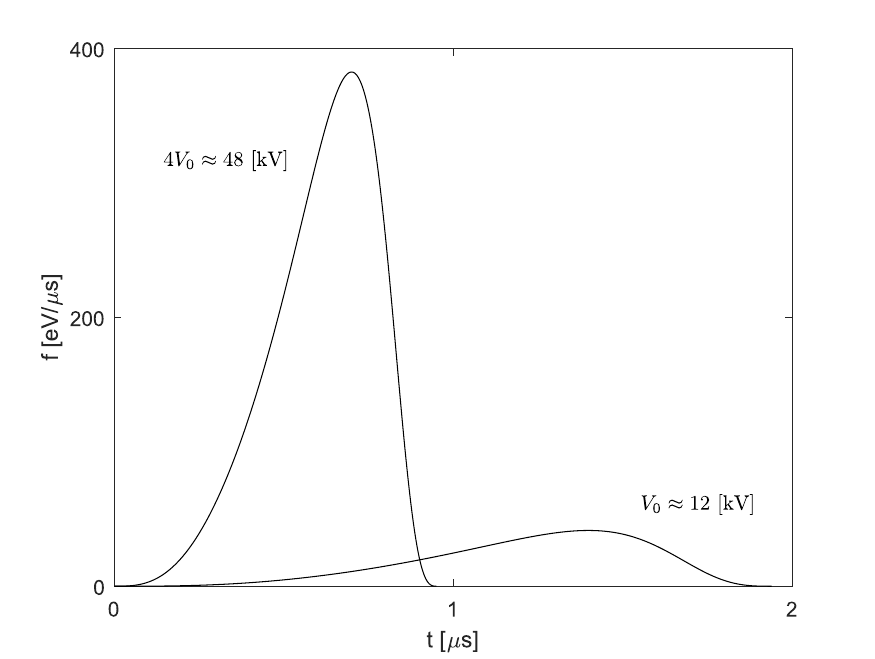}
\caption{The function $f=(2/3)m\left[(-1/r)\left(dr/dt\right)^3\right]$ versus time $t$ for the case $V_0\approx 12$~[kV] and the case $4V_0\approx 48$~[kV].}
\end{figure}

We went a step further by performing targeted simulations. Then, based on that bunch of information, we discovered heuristically that the temperature of the plasma at pinching roughly scales linearly with the initial voltage at the bank of capacitors.
In the present context, that translate into the formula
 
\begin{equation}
\left(\frac{k_BT}{k_BT_0}\right)=\lambda.
\end{equation} 
On the other hand, Equation (41) states $(E/E_0)=\lambda^2$, hence the scaling rule

\begin{equation}
\left(\frac{k_BT}{k_BT_0}\right)=\left(\frac{E}{E_0}\right)^{\frac{1}{2}}   
\end{equation}
follows immediately.
Incidentally, let us take this opportunity to mention that along all over the simulations that lead to this result, the values of the efficiencies: $\eta_{pl}\approx 51\,\%$, $\eta_{th}\approx 68\,\%$ and, of course, $\eta \approx 35\,\%$ prevail.

What Equation (47) tells us is that by simply increasing $V_0$ sufficiently, the plasma might reach temperatures within the range of thermonuclear interest. As an example, let us return to our reference experiment but increase its voltage by a factor 
$\lambda=30$, \textit{i.e.} the voltage at the bank of capacitors is now approximately $360\,[kV]$ and the electric field applied to the gas is about $10\, [kV/cm]$. The corresponding simulation indicates that the temperature of  plasma at pinching rises from the value $k_BT_0\approx 32\,eV$ to the value $k_BT\approx 1.1\, keV$.

This result is however disturbing and the only way to check it is through experiments. Indeed, when the voltage at the bank of capacitors is very high, the dynamics of the current sheath is too rapid, so that the randomization of plasma particles motion is unlikely. In this way, there would be neither enough production of internal energy of the plasma nor sufficient increase of the temperature of the plasma, so that the Z pinch device would behave simply as a particle accelerator. It is worth emphasizing that if that were the case, the MSP equations - based on the assumption that the internal energy of the plasma manifests itself as kinetic pressure outward on the plasma sheath- would no longer be valid. Instead, the appropriate framework for dealing with this case falls within the scope of theories of shock heating \cite{hammer}.

Whatever it is, let us consider for a moment a counterfactual. This is, according to the MSP equations, our experimental setup powered by a bank of capacitors charged to $360\,[kV]$ is capable of bringing the plasma temperature to $k_BT\approx 1.1\, keV$. Although this temperature is far from the temperature necessary for thermonuclear fusion, it is still interesting because it would give rise to a profuse emission of neutrons. Then, the problem arises that our experimental setup is too small for producing a relevant amount of fusion energy. Thus, to mantain the temperature achieved in that experiment and simultaneously produce a significant amount of fusion energy, we would have to enlarge proportionally by a factor, say $\xi$, the spatial and electrical specifications of the experimental setup. Therefore, finally the voltage across the bank of capacitors would be equal to $\xi\times 360\,[kV]$. Then, the issue is that 
$\xi\times 360\,[kV]$ may become too high a voltage and hence only actual practice can demonstrate whether such an experimental setup is feasible or not.

\section{Conclusions}

In the framework of the snowplow model where ohmic resistances and hydromagnetic instabilities are ignored we have obtained that 

\begin{enumerate}
\item For simulations that replicate successful  real-life experiments, the efficiency at which the electrostatic energy stored in the bank of capacitors transforms to internal energy of the plasma is approximately $35\,\%$.
\item While the magnetic field is efficient at confining the plasma it is not so efficient at transferring mechanical energy to the plasma.
\item  This result would be generalizable to all experimental setups where the snowplow model is applicable, \textit{e.g.} plasma focus, simple torus, etc.
\item For a given experimental setup, the plasma temperature at pinching increases linearly with the voltage applied to the plasma.
\item As far as our current knowledge is corcerned, the scaling law for discharges in Z pinch devices can be summarized as

\begin{equation}
\left(\frac{k_BT}{k_BT_0}\right)=\left(\frac{E}{E_0}\right)^n
\end{equation}
where depending on the case, the exponent $n$ takes one of the following three values:

$n=0$ if the parameters $r_0$, $l_0$, $C_0$, $L_0$ and $V_0$ are scaled all them to the same extent; $n=1/3$ if only the paramenters $r_0$, $C_0$ and $V_0$ are scaled to the same extent; $n=1/2$ if only the voltage $V_0$ is scaled.
\end{enumerate}


\begin{thebibliography}{99}
\bibitem{gamow} G. Gamow and C. L. Critchfield, 'Theory of Atomic Nucleus and Nuclear Energy-Sources', Oxford University Press, at the Clarendon Press, Oxford 1949.
\bibitem{teller}Edward Teller, Peaceful Uses of Fusion, Report UCRL-5257 Rev. University of California Radiation Laboratory, Livermore, California, July 3, 1958.
\bibitem{mccall2} Gene H. McCall, Cloud and Microjet Mix: A possible Source of Yield limitation of the National Ignition Facility Targets, arXiv.org, 28 Aug 2022, [arXiv:2208.13131v1].
\bibitem{bennett}W. H. Bennett: Magnetically Self-focusing Sreams, Phys. Rev. \textbf{45}, 890 (1934).
\bibitem{tonks}L. Tonks, 'Theory of Magnetic Effects in the Plasma of an Arc', Phys. Rev. \textbf{56}, 360 (1939). 
\bibitem{jackson}J. D. Jackson, 'Classical Electrodynamics', John Wiley and Sons, New York (1975).
\bibitem{reitz-milford}John R. Reitz and Frederick J. Milford, 'Foundations of Electromagnetic Theory', Addison-Wesley Publishing Company, Inc. Reading Massachusetts, U. S. A. (1960).
\bibitem{krall}N. A. Krall and A. W. Trivelpiece, 'Principles of Plasma Physics', McGraw-Hill, Inc., New York (1973).
\bibitem{glasstone}S. Glasstone and R. H. Lovberg: 'Controlled Thermonuclear Reactions'
, D. Van Nostrand Company, Princeton, N. J. (1960).
\bibitem{bishop}Amasa S. Bishop, 'Project Sherwood- The U. S. Program in Controlled Fusion', Addison-Wesley, Reading, Massachusetts (1958).
\bibitem{hagler}M. O. Hagler and M. Kristiansen, 'An Introduction to Controlled Thermonuclear Fusion', Lexington Books D. C. Heath and Company Lexington Massachusetts Toronto (1977).
\bibitem{post}Richard F. Post, Controlled Fusion Research- An Appplication of the Physics of High Temperature Plasmas, Rev. Mod. Phys. \textbf{28}, 338 (1956).
\bibitem{kolb}Alan C. Kolb, Magnetic Compression of Plasmas, Rev. Mod. Phys. \textbf{32}, 748 (1960).
\bibitem{ryutov}D. D. Ryutov, M. S. Derzon and M. K. Matzen: The physics of fast Z pinches, Rev. Mod. Phys. \textbf{72}, 167 (2000).
\bibitem{mather}J. Mather: An Intense Source of Neutrons from the Dense Plasma Focus, Intense Neutron Sources: Proceedings of a United States Atomic Energy Commission/European Nuclear Energy Agency seminar, Santa Fe, New Mexico, 19-23 September 1966.
\bibitem{rosenbluth}M. N. Rosenbluth, R. Garwin and A. Rosenbluth: Infinite Conductivity Theory of the Pinch, Report LA-1850, Los Alamos Scientific Laboratory, New Mexico, September 1954.
\bibitem{cardenas}Miguel Cárdenas, Alejandro Nettle and Leandro Núñez, Scaling Law for Discharges in Z pinch Devices, arXiv.org, 12 February 2025, [arXiv:2502.08570v1].
\bibitem{lawson}J. D. Lawson: Some Criteria for a Power Producing Thermonuclear Reactor, Proc. Phys. Soc. B, \textbf{70} (1957).
\bibitem{meek}J. M. Meek and J. D. Craggs, 'Electrical Breakdown of Gases', Oxford University Press, at the Clarendon Press, Oxford (1953).
\bibitem{spitzer}Lyman Spitzer, Jr., 'Physics of Fully Ionized Gases', Interscience Publishers, Inc., New York (1956).
\bibitem{anderson}Oscar A. Anderson, William R. Baker,Stirling A. Colgate, John Ise, Jr., and Robert V. Pyle, Neutron Production in Linear Deuterium Pinches, Phys. Rev. \textbf{110}, 1375 (1958).
\bibitem{hai}M. J. Bernstein and F. Hai, Evidence for Nonthermonuclear Neutron Production in a Plasma Focus Discharge, Phys. Lett. \textbf{31A}, 317 (1970).
\bibitem{mc call}Gene H. McCall, Calculation of Neutron Yield from a Dense Z Pinch, Phys. Rev. Lett. \textbf{62}, 1986 (1989).
\bibitem{hammer}Alejandro Mesa Dame, Eric S. Lavine and David A. Hammer, A Comprehensive Analytical Model of the Dynamics Z-Pinch, arXiv.org, 23 May 2025, [arXiv:2505.18067v1]. 
\end{thebibliography}
\end{document}